\documentclass[12pt]{article}
\usepackage{xcolor}
\usepackage{hyperref}

\usepackage{amssymb}
\usepackage{amsmath}
\usepackage{amsfonts}
\usepackage{graphicx}
\usepackage{mathrsfs}
\usepackage{comment}
\usepackage{cite}
\usepackage{authblk}
\usepackage{array}
\usepackage{subcaption}

\newcommand{\Slash}[1]{{\ooalign{\hfil#1\hfil\crcr\raise.167ex\hbox{/}}}}

\newcommand{\beq}{\begin{equation}}  \newcommand{\eeq}{\end{equation}}
\newcommand{\bef}{\begin{figure}}  \newcommand{\eef}{\end{figure}}
\newcommand{\bec}{\begin{center}}  \newcommand{\eec}{\end{center}}
\newcommand{\non}{\nonumber}  
\newcommand{\laq}[1]{\label{eq:#1}}  

\newcommand{\Eq}[1]{Eq.(\ref{eq:#1})}

\def\({\left(}
\def\){\right)}

\def\O{\mathcal{O}}

\newcommand{\AND}{~{\rm and}~}
\newcommand{\EV}{ {\rm \, eV} }

\def\d{\delta}
\def\e{\epsilon}

\def\g{\gamma}

\def\m{\mu}
\def\n{\nu}

\def\*{\dagger}

\usepackage[normalem]{ulem}

\begin{document}
\begin{titlepage}
\begin{center}

\vspace{1.0cm}

{\Large\bf 
Novel Limits on Dark Photon Mixing from Radiation Safety
}

\vspace{1.0cm}

{\bf  Wen Yin}

\vspace{1.0cm}
{\em 
$^{1}${Department of Physics, Tokyo Metropolitan University, Tokyo 192-0397, Japan\\}
}

\vspace{1.0cm}
\abstract{
I propose a novel laboratory search for dark photons based on radiation-safety monitoring at synchrotron radiation facilities, including NanoTerasu, SPring-8, KEK-PF, and ESRF.  
Dark photons can be produced parasitically in undulators or via photon-mirror interactions, and subsequently traverse optical systems and shielding.  
Taking into account quantum effects and the internal structure of undulators, mirrors, and detectors, I show that even a simple Geiger-M\"uller counter, routinely used for radiation-safety monitoring, can detect such dark photons outside the shielding and set competitive limits on the kinetic mixing parameter down to $\chi \lesssim 5\times 10^{-6}$ in the eV mass range, providing some of the strongest bounds among laboratory searches.  
Because radiation safety is strictly regulated, the resulting limits can be regarded as robust and realistic constraints. 
}

\end{center}
\end{titlepage}

\setcounter{footnote}{0}
\section{Introduction}

The search for light and weakly interacting particles, such as axions and dark photons, has attracted significant attention since they provide simple extensions of the Standard Model and can explain various aspects such as dark matter, inflation and the smallness of the nucleon electric dipole moment.  
There are two broad approaches to searching for such particles: one may attempt to detect them as they are produced in nature---for instance, from astrophysical or cosmological sources---or to generate and detect them under controlled laboratory conditions. The latter approach offers the advantage that systematic uncertainties can be more reliably understood and managed, thereby allowing for more robust searches, although the sensitivity is usually lower compared to astrophysical observations.  


Among laboratory searches, light-shining-through-a-wall (LSW) experiments represent a well-established and versatile method~\cite{Redondo:2010dp,Ehret:2010mh,Inada:2013tx,Betz:2013dza,Halliday:2024lca,Battesti:2010dm,DellaValle:2015xxa,Inada:2016jzh,OSQAR:2015qdv,Sikivie:2007qm,Carenza:2025uwx}.  
In these setups, an intense photon beam is directed at an opaque barrier, and hypothetical particles such as axion-like particles (ALPs) or dark photons may be produced through photon-new particle conversion. These particles, unlike photons, can traverse the barrier and subsequently reconvert into photons that can be detected. Dedicated LSW experiments have been implemented at laser facilities and synchrotron light sources.

Recently, it has been proposed by the present author and J.~Yoshida in \cite{Yin:2024rjb} that synchrotron radiation facilities, such as NanoTerasu, SPring-8, the KEK Photon Factory (PF), and the European Synchrotron Radiation Facility (ESRF), allow for a novel ``parasitic''  LSW-type search.  
The key idea is that ALPs or dark photons may already be generated in the undulators of the light source.  
Thus, by placing a detector outside the shielding along the beam line, one can exploit the intense synchrotron photon flux without interfering with the primary operation of the facility.  Indeed, such positions usually exist. 
In \cite{Yin:2025awb}, the dark photon system for the parasitic LSW is carefully studied by taking into account the momentum integration within the wave packet, the attenuation of photons in a lead wall, and the suppression of mixing due to the medium effects.  
There, the detector is assumed to be an ideal one.  
It was shown that, depending on the mass range, the conversion rate can differ significantly from the na\"{i}ve application of the well-known photon-dark photon oscillation formula. 

In this paper, I study a more realistic setup inspired by various synchrotron facility designs, and provide a detailed investigation of a simple and well-known detector: the Geiger-M\"uller (GM) counter.  
In particular, I demonstrate that even with such a basic radiation monitor---specifically, a GM counter that can serve as an efficient detector for X-ray dark photons---one can already place competitive bounds on dark photons in the mass range ${\,\mathrm{meV}}$--${10\,\mathrm{eV}}$.\footnote{See also a variety of models and searches that motivate dark photons in this mass range for dark matter production~\cite{Moroi:2020has,Moroi:2020bkq,Nakayama:2021avl,Yin:2023jjj,Sakurai:2024apm,Fujita:2023axo,Kitajima:2023fun,Graham:2015rva} and for addressing the Hubble tension~\cite{Nakagawa:2022knn}.  
In particular, several dark matter production mechanisms point to this mass window~\cite{Yin:2023jjj,Sakurai:2024apm}, given that the matter-radiation equality occurs around the eV temperature scale, similar to the prediction of the hot dark matter paradigm.  
See also proposals for dark matter decay searches in this mass range~\cite{Bessho:2022yyu,Bessho:2024tpl}.  
For dark matter searches, one can consider $\gamma' \to \gamma \phi$ with $\phi$ being a light scalar field, producing a narrow line.  
Furthermore, multi-laser collision experiments may probe dark photons in this mass range via the $\gamma'\gamma\gamma\gamma$ coupling, similar to the ALP search proposed in Ref.~\cite{Nobuhiro:2020fub}.}

\begin{figure}[t!]
  \begin{center}  
\includegraphics[width=85mm]{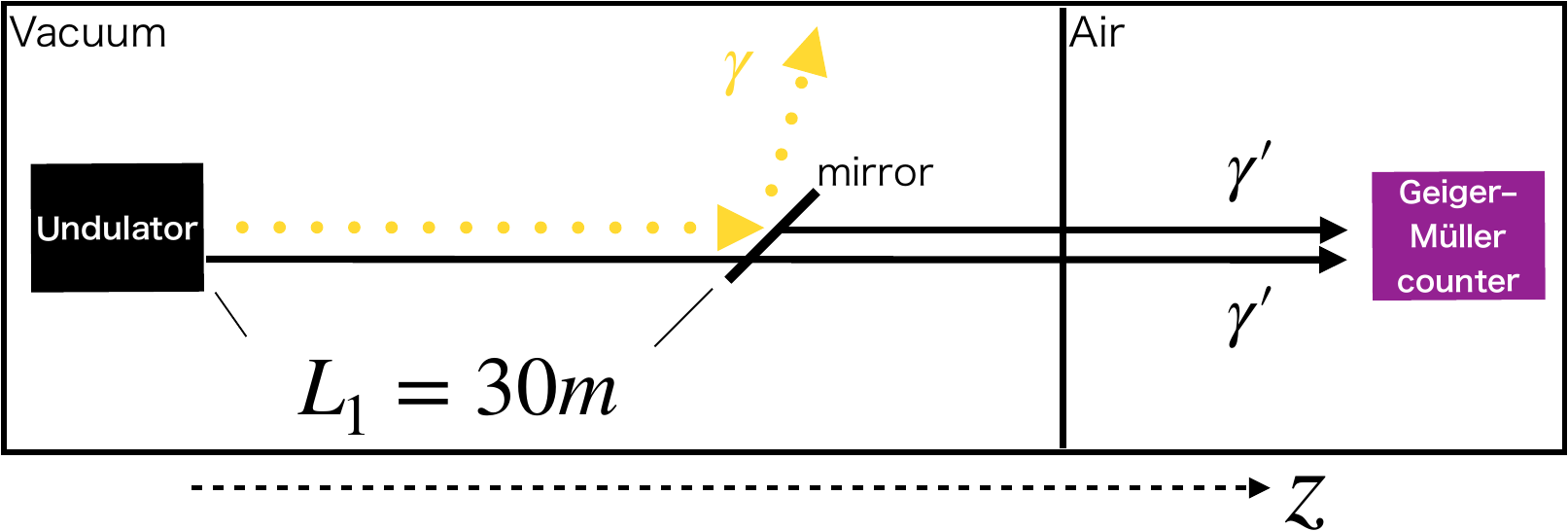}\hspace{1cm}
\includegraphics[width=40mm]{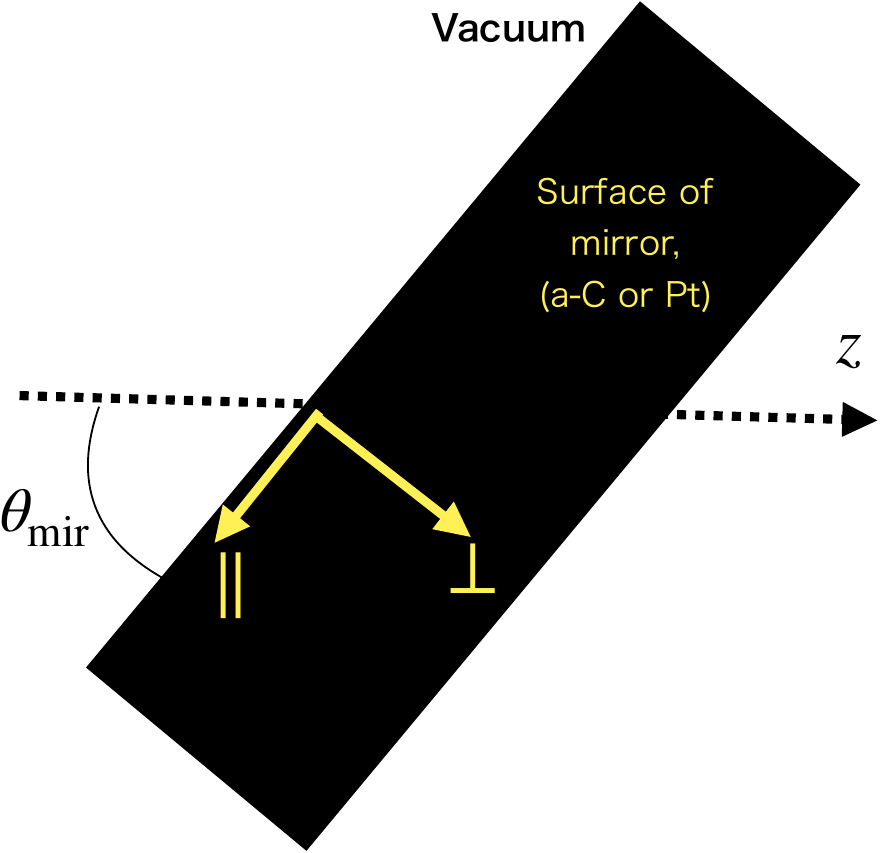}
  \end{center}
    \caption{The schematic view of the setup under consideration (left panel).  
Throughout this paper I take $L_1 = 30$\,m.  
The mirror is aligned so that most photon-like states are directed into the experimental hutches.  
A magnified view of the mirror is shown in the right panel, where the axes relevant for studying the transmission of the dark-photon-like state are indicated.    }
    \label{fig:setup}
    \end{figure}

The experimental setup is shown in the left panel of Fig.~\ref{fig:setup}.
Before going into details, let me provide an order-of-magnitude estimate to illustrate the impact of the radiation safety limit.  
First, the natural radioactivity background in dose is of order
\beq
D_{\rm natural} \lesssim 0.1\,\mu \mathrm{Sv}\cdot \mathrm{h}^{-1}.
\eeq
For a typical GM counter calibrated with Cs$^{137}$ (662 keV), a dose rate of 
$0.1 \,\mu$Sv/h corresponds to about 10 counts per minute.  
This conversion factor is detector-dependent, but we adopt it here as a reasonable representative value.  Note that the GM counter measures only the photon number flux, while the resulting dose rate is quoted assuming the $662$\,keV $\gamma$-ray line of Cs${}^{137}$.

As an example, consider an undulator producing photons in the keV range with a photon beam power of ${\rm 3\,kW}$.  
The photon production rate can be estimated as
\beq
\laq{photo}
\dot{n}_\gamma 
\sim 10^{21}\,\mathrm{min}^{-1}.
\eeq
In what follows, I neglect the dependence on beamline energy and photon beam power, and adopt this value as a representative choice for this work (note that for certain beamlines the photon can be as large as $\O(10)\,\mathrm{kW}$). 
Since both the production and detection rates for dark photon scale as $\chi^2$, requiring that the GM counter does not exceeds the natural radioactivity background significantly, one obtains the very na\"{i}ve estimate
\beq
\chi_{\rm limit}^4 \times 10^{21}/{\rm min} \sim 10/{\rm min}
\quad \Rightarrow \quad \chi_{\rm limit} \sim 10^{-5}.
\eeq
This limit is already comparable to existing laboratory bounds, $10^{-5}\text{--}10^{-4}$, obtained using the SPring-8 beamline~\cite{Inada:2013tx}. 
The reason is that the available photon flux directly at the undulator is many orders of magnitude larger than that typically accessible inside an experimental hutch, where optical components such as mirrors, filters, and spectrometers strongly reduce the flux. 
For instance, the mirrors in Fig.~\ref{fig:setup} are used to transport photons into the experimental hutches. 
In contrast, I consider dark photons converted from the primary photon beam.

Next, I will carefully study the constraints from radiation safety and show that, depending on the parameter region, one may obtain much stronger or weaker limits due to the detector response to dark photons and due to the material and quantum effects.

\section{Setup and production from undulator}

We assume a minimal dark photon $A'_\mu$ with mass $m$ and kinetic mixing~$\chi$ with the visible photon~$A_\mu$. The Lagrangian in the interaction basis is given by
\begin{equation}
\mathcal L \;=\; -\frac14 F_{\mu\nu}F^{\mu\nu}
                 -\frac14 F'_{\mu\nu}F'^{\mu\nu}
                 +\frac{\chi}{2}\,F_{\mu\nu}F'^{\mu\nu}
                 +\frac12 m^2 A'_\mu A'^{\mu}
                 - J^\mu A_\mu .
\label{eq:L}
\end{equation}
Here $F_{\m\n},F'_{\m\n}$ denote the field strength of $A_\m, \AND A'_\m,$ respectively. 
In the interaction basis, only the visible photon field $A_\mu$ couples to the electromagnetic current $J^\mu$ of ordinary charged particles. 
The dark photon $A'_\mu$ interacts with matter only through the kinetic mixing term, and can thus be produced when a high-intensity photon flux is available, such as from an undulator radiation source.

An \emph{undulator} is a periodic magnetic structure widely used in synchrotron and free-electron-laser facilities. 
When a relativistic electron beam traverses an array of alternating-polarity magnets, it follows a sinusoidal trajectory and emits synchrotron radiation. 
Because the motion is periodic, radiation from successive oscillations interferes constructively, resulting in a highly brilliant, collimated, and quasi-monochromatic X-ray beam with a narrow spectral bandwidth.

The undulator is characterized by parameters $k$, $K$, $\phi$, and $\kappa$ in the notation of Refs.\,\cite{Yin:2024rjb, Yin:2025awb}. 
Here $k$ denotes the undulator wave number, related to the period $\lambda_u$ by $k = 2\pi/\lambda_u$. 
The parameter $K$ represents the strength of the electron wiggling, while $\kappa$ and $\phi$ encode the undulator geometry. 
For definiteness, I set $\kappa=1$ and $\phi=0$, corresponding to a helical undulator. 
Varying these parameters does not drastically affect the sensitivity limits discussed in this work. 

The number of photons produced per electron traversing the undulator is
\begin{equation}
N_{\gamma}=\sum_{\epsilon^{\pm}} \int \frac{d^3 \vec k_{\gamma}}{(2\pi)^3} 
\left|f_{\gamma}(\vec k_{\gamma})\right|^2 . \laq{photonum}
\end{equation}
$\e_\pm$ denotes the two photon polarizations. 
The photon wave packet $f_{\gamma}(\vec k_{\gamma})$, neglecting mixing, follows from standard calculations and is well known (see the Appendix for an estimate)~\cite{Jackson:1998nia,Yin:2024rjb,Yin:2025awb}.  
I will retain the $\kappa$ and $\phi$ dependence in the formulas, and analyses with different undulator configurations are straightforward. I also only focus on the fundamental harmonic for $f_\g$ in this study for simplicity.

By neglecting matter effects, we introduce the mass eigenfields $A_\m^{(1)}$ and $A_\m^{(2)}$, in which both the kinetic and mass terms are diagonalized. 
Here $A_\m^{(1)}$ corresponds to the massless photon-like field, and $A_\m^{(2)}$ to the massive dark-photon-like field. 
In this basis, the electromagnetic current couples as
\begin{equation}
J^\mu A_\mu \;\simeq\; J^\mu \left(A^{(1)}_\mu -\chi\, A^{(2)}_\mu \right),
\end{equation}
neglecting $\O(\chi^2)$ terms. 

When the dark photon mass is negligible, one obtains
\begin{equation}
f^{(1)} \;\approx\; f_{\gamma}, 
\qquad 
f^{(2)} \;\approx\;- \chi\, f_{\gamma}. \laq{amp}
\end{equation}
Importantly~\cite{Yin:2025awb},
\begin{equation}
f^{(2)} \;\to\; 0 \quad \text{when}\quad 
m > m_{1}\equiv \frac{k\gamma}{\sqrt{1+\tfrac{1+\kappa^2}{2}K^2}}.
\end{equation}
This suppression arises because synchrotron radiation in an undulator can be understood as boosted dipole radiation of the electron. 
In the electron rest frame, the oscillation frequency is 
$\tfrac{k\gamma}{\sqrt{1+\tfrac{1+\kappa^2}{2}K^2}}$, 
so a dark photon heavier than this value cannot be efficiently produced. 
For the numerical simulations, I approximate this suppression by multiplying $f^{(2)}$ with a Heaviside step function that enforces the above inequality, which provides a good approximation to the full calculation~\cite{Yin:2025awb}.

Throughout this paper, I neglect the longitudinal mode, whose production and detection are both suppressed by $m^2/\omega^2$ and is therefore irrelevant in the regime $m\ll \omega$ considered here.

\section{Propagation across the mirror}

I now discuss the propagation of the photon--dark photon system when the synchrotron beam encounters the first mirror. 
In practice, the mirror reflects or absorbs most of the photon-like state, while the weakly interacting dark-photon-like state can propagate essentially unimpeded through the mirror material. 
The aim here is to describe how the photon/dark photon system evolves in the presence of the mirror.  
To this end, I assume that the mirror coating is sufficiently thick that the photon-like state at a shallow incidence angle cannot penetrate. 
Under these assumptions, the problem reduces to a simple boundary-value problem at the vacuum-mirror interface. 

For later convenience, I define (see the right panel of Fig.\ref{fig:setup})
\beq
\theta_{\rm mir} \;\equiv\; \frac{\pi}{2}- \arccos\!\left(\hat{n}\cdot \hat{e}_z\right),
\eeq
where $\hat n$ is the normal vector of the mirror and $\hat{e}_z$ is the unit vector along the $z$-direction (the beam axis).  
Since the photon (dark photon) beam spread is of order $1/\gamma \ll 10^{-4}$ (see Table.\ref{tab:1}), which is much smaller than a typical mirror angle $\theta_{\rm mir}\gtrsim 0.1^\circ$, I neglect the angular dependence when estimating the transmission.

I again work in the interaction basis while including medium effects through the self-energy. 
Restricting to transverse modes, the linearized equations of motion in a homogeneous, isotropic medium read
\begin{align}
\non \Big[\omega^2 + \partial_{\perp}^2 - \vec k_\parallel^2 - \Pi_{i}(\omega)\Big] A_\mu
\;+\; \chi\, m^2 A'_\mu &= 0, \\
\Big[\omega^2 + \partial_\perp^2 - \vec k_\parallel^2 - m^2\Big] A'_\mu
\;+\; \chi\, m^2 A_\mu &= 0,
\label{eq:EOM-matrix}
\end{align}
where we factorize the dependence 
$\exp(i\vec{k}_\parallel \cdot \vec{x}_\parallel - i\omega t)$, and the self-energy $\Pi(\omega)$ is taken to be complex in order to encode absorption. 
Here $\parallel$ ($\perp$) denote the momentum components parallel (perpendicular) to the mirror surface (see the right panel of Fig.~\ref{fig:setup}). According to the boundary problem, $\vec{k}_\parallel$ and $\omega$ are taken in $i=\mathrm{vac}$ (vacuum) and $i=\mathrm{mir}$ (mirror material).

For an isotropic dielectric with refractive index $n(\omega)$, one may write
\begin{equation}
\Pi_{i}(\omega) = \omega^2\!\left[\,1 - n_i^2(\omega)\,\right].
\end{equation}
Thus $\Pi_{\rm vac}=0$, while in the mirror region one generally has a complex refractive index
\begin{equation}
n_{\rm mir}(\omega) = 1 - \delta + i\beta ,
\end{equation}
where $\delta$ encodes dispersion and $\beta$ encodes absorption. 
Values of $\delta$ and $\beta$ for relevant materials can be obtained from standard references such as Henke et al.~\cite{henke1993x}\footnote{\url{https://henke.lbl.gov/optical_constants/}}, and are illustrated in Fig.~\ref{fig:n}.

In a given homogeneous region~$i$, the $2\times 2$ system \eqref{eq:EOM-matrix} can be diagonalized by the effective mixing angle (to leading order in $\chi$),
\begin{equation}
\chi^{\rm eff}_i(\omega) \;=\; \frac{\chi m^2}{\,m^2-\Pi_{i}(\omega)\,}.
\label{eq:theta-eff}
\end{equation}
The corresponding propagation eigenfields are
\begin{equation}
A_{i}^{(1)\mu} \;=\; A^\mu + \chi^{\rm eff}_i \,A'^\mu,
\qquad
A_{i}^{(2)\mu} \;=\; -\chi^{\rm eff}_i A^\mu + A'^\mu,
\label{eq:eigen}
\end{equation}
with dispersion relations
\begin{align}
k_{\perp,i}^{(1)}(\omega,k_\parallel)
&= \sqrt{\,n_i^2\omega^2 - k_\parallel^2\,} + \mathcal{O}(\chi^2), \\
k_{\perp,i}^{(2)}(\omega,k_\parallel)
&= \sqrt{\,\omega^2 - m^2 - k_\parallel^2\,} + \mathcal{O}(\chi^2).
\label{eq:dispersions}
\end{align}
When ${\rm Re}(n_{\rm mir}) < 1$ i.e. $\d_{\rm mir} > 0$, one can see that for sufficiently large $k_{\parallel}$, or small $\theta_{\rm mir}$, the quantity $k_{\perp,\rm mir}^{(1)}$ becomes imaginary. This condition is easily estimated to be $\theta_{\rm mir}<\sqrt{2\delta}$.  
In this regime the modes are strongly attenuated inside the mirror, leading to nearly perfect reflection, which is precisely the operating principle of an X-ray mirror.

To estimate the transition probability, the continuity condition for the fields must be satisfied. 
This condition applies to the interaction-basis fields,\footnote{There is a polarization dependence in the continuity condition due to the medium-dependent Coulomb gauge. 
I neglect this effect, as it hardly impacts the transmission considered here, given that $n_{\alpha}\approx 1$ in both the mirror and vacuum in the X-ray region.} namely at the mirror interface:
\begin{equation}
A^{(')\mu}_{\rm vac} = A^{(')\mu}_{\rm mir}, 
\qquad 
\partial_{\perp} A^{(')\mu}_{\rm vac} = \partial_{\perp} A^{(')\mu}_{\rm mir}.
\label{cont}
\end{equation}
Note that the equations must be expanded in terms of the mass eigenfunctions for the Fourier modes satisfying the dispersion relation. 
In this way, taking account reflection, transmission components for each mass eigen modes, I obtain
\begin{equation}
\tilde A_{\rm mir}^{(2)\mu} \;\approx\; 
\frac{k_{\perp,\rm vac}^{(1)} \big(k_{\perp,\rm mir}^{(1)}+k_{\perp,\rm vac}^{(2)}\big)}
     {k_{\perp,\rm vac}^{(2)} \big(k_{\perp,\rm mir}^{(1)}+k_{\perp,\rm vac}^{(1)}\big)} 
\,\Delta\chi\, \tilde A_{\rm vac}^{(1)\mu}
\;+\; \tilde A_{\rm vac}^{(2)\mu},
\laq{match}
\end{equation}
where
\begin{equation}
\Delta\chi \;=\; \chi^{\rm eff}_{\rm mir}-\chi^{\rm eff}_{\rm vac},
\end{equation}
and $\tilde A_{i}^{(\alpha)\mu}$ denote the Fourier modes satisfying the dispersion relation for the corresponding fields.
Here $\tilde{A}^{(1)\mu}_{\rm vac}=\tilde{A}^{(1)\mu}_{\rm vac}(\vec k^{(1)}_{\rm vac})$ and 
$\tilde{A}^{(2)\mu}_{\rm vac}=\tilde{A}^{(2)\mu}_{\rm vac}(\vec k^{(2)}_{\rm vac})$.  
The momenta $\vec k^{(\alpha)}_{\rm vac}$ are functions of $\vec k^{(2)}_{\rm mir}$ satisfying 
\begin{align}
\label{boundary1}
\omega = \sqrt{ (k^{(1)}_{\perp,\rm vac})^2+\vec k_{\parallel}^2 }
&= \sqrt{ (k^{(2)}_{\perp,\rm mir})^2+\vec k_{\parallel}^2+m^2},\\
\label{boundary2}
\vec k^{(2)}_{\rm mir}&=\vec k^{(2)}_{\rm vac}.
\end{align}

From this one finds that when
\begin{equation}
m > m_{2} \;\equiv\; \sqrt{\omega^2-\vec{k}_{\parallel}^2},
\end{equation}
the quantity $k_{\perp,\rm mir}^{(2)}$ becomes imaginary, and the dark-photon-like initial state produced from the photon-like state does not propagate in the mirror. 
Since I assume the mirror is sufficiently thick, I do not consider penetration of the dark photon in this region. This condition does not apply to dark photons directly produced from the undulator, because the momentum above does not satisfy the dispersion relation for the initial state. 

When $m\sim m_{2}$, the direction of the dark photon is slightly shifted from the beam axis. 
For 
\begin{equation}
m \ll m_{3}, \qquad 
m_{3} \;\equiv\; |\vec{k}_{\perp,\rm mir}^{(2)}|\simeq \sin(\theta_{\rm mir})\,|\vec{k}_{\rm mir}^{(2)}|,
\end{equation}
the contribution of the mass to the propagation angle can be neglected. 

I arrive at the number of massive states that pass through the mirror,
\begin{align}
N_{\rm mir}^{(2)}&=\sum_{\e^{\pm}}\int_{z=L_1+\e} dt\, d^2x_\parallel  
  \big(A^{(2)\mu}_{\rm mir}(\vec x,t)\big)^* \big(-i \partial_{\perp}\big) 
  A^{(2)}_{\mu,\,\rm mir}(\vec x,t) + \mathrm{h.c.} \\
&=\sum_{\e^{\pm}}\int \frac{d^3 k^{(2)}_{\rm mir}}{(2\pi)^3} 
  2\,(\tilde A^{(2)\mu}_{\rm mir})^* \,\omega\, \tilde A^{(2)}_{\mu,\,\rm mir} \\
&=\sum_{\e^{\pm}}\int \frac{d^3 k^{(2)}_{\rm mir}}{(2\pi)^3} 
  2\omega \left|-\frac{k_{\perp,\rm vac}^{(1)} 
  \big(k_{\perp,\rm mir}^{(1)}+k_{\perp,\rm vac}^{(2)}\big)}
  {k_{\perp,\rm vac}^{(2)} \big(k_{\perp,\rm mir}^{(1)}+k_{\perp,\rm vac}^{(1)}\big)} 
  \,\Delta \chi^{\rm eff} \,\tilde A^{(1)\mu}_{\rm vac}
  + \tilde A^{(2)\mu}_{\rm vac}\right|^2 .
\end{align}
I have performed the integrals over $t,x,y$ and the momentum integral in $A^{(2)}_{\mu,\,\rm mir}$. In the last formula, all the momenta satisfy the previous boundary condition and expressed in terms of $\vec k_{\rm mir}^{(2)}.$

By using the circularly polarized amplitudes, I then obtain\footnote{One can easily see that $|{\cal M}|^2$ with small $m$ reduces to the usual oscillation form multiplied by the wave packet when we take $\chi^{\rm eff}_{\rm mir}=0$, corresponding to ${\rm Im}\,\Pi_{\rm mir}\gg m^2$, i.e. when the mirror effectively acts as a perfect ``wall.'' }
\begin{align}
N_{\rm mir}^{(2)}
&\simeq \sum_{\e^{\pm}}\int \frac{d^3 k^{(2)}_{\rm mir}}{(2\pi)^3}  
\left|{\cal M}\right|^2 ,\\
{\cal M}&
\approx \(-\frac{k_{\perp,\rm vac}^{(1)} \big(k_{\perp,\rm mir}^{(1)}+k_{\perp,\rm vac}^{(2)}\big)}
{k_{\perp,\rm vac}^{(2)} \big(k_{\perp,\rm mir}^{(1)}+k_{\perp,\rm vac}^{(1)}\big)} 
\Delta \chi^{\rm eff} \Theta(m-m_{3})
+ e^{-i \tfrac{m^2}{2\omega}L_1} \chi \Theta(m-m_{1})\)f_{\g}
\end{align}
where I used the wave packets discussed in \Eq{amp}, $L_1$ denotes the distance between the undulator and the mirror, and I have expanded $m^2$ in the phase factor.  
Heaviside step functions, $\Theta$, are introduced to account for suppression in the production and attenuation inside the mirror.  
The factor $2\omega$ cancels due to the definition of $f_\g$.  
Note that the decoherence effect is automatically incorporated in this form because the calculation is performed with wave packets (see Ref.~\cite{Yin:2025awb}).  

This expression gives the number of dark-photon-like mass eigenstates propagating across the mirror.  
Afterwards, the state can pass through all optical systems and shielding, owing to its small coupling proportional to the mixing parameter.  

In general, further photon production can also occur: when the heavier mass eigenstate propagates through the mirror, it can regenerate lighter mass eigenstates on the other side.  
Moreover, realistic mirrors often consist of several layers, within which the photon-like state gradually attenuates, and at each boundary additional dark-photon-like states may be produced.  
Similar effects can arise in other optical components and even in shielding.  
In this work, however, I ignore such contributions, which would only enhance the sensitivity, and instead focus on the dark-photon-like heavy states produced directly in the undulator and at the first interaction with the material on the mirror surface such as Amorphous carbon or Pt.

\section{Novel limits from radiation safety}

Once the dark-photon-like states are produced, most of them pass through all optical systems, shielding, and even air. However, due to their non-vanishing mixing, there remains a rare probability for them to interact with a ``photon detector.'' In this section, I discuss their detection and show that the simple requirement of radiation safety at the facility can already place stringent limits on the dark photon mixing. The safety is legally regulated and strictly monitored, particularly at synchrotron radiation facilities.

As a concrete example, let us assume an argon (Ar)-filled gas detector (see \cite{Knoll:2010xta} for details).  
It consists of a metal cylinder (length $\sim$10 cm, diameter $\sim$1 cm) serving as the cathode, and a thin ($\sim$50\,$\mu$m) central anode wire.  
By contrast, the beam spread is of order $1/\gamma \lesssim 10^{-4}$. Thus the GM counter can cover the entire dark-photon-like beam, provided the counter is placed within a distance of $\gamma \times 1\,$cm $\gtrsim 100$\,m from the undulator.  

The chamber is filled with high-purity argon at $\sim$1 atm, together with a small fraction (5--20\%) of quench gas (e.g.\,CH$_4$ or alcohol vapor).  
Incident radiation or wall-converted photoelectrons produce primary Ar ion pairs along track lengths of a few millimeters to centimeters. A high voltage (several hundred to a few thousand volts) between cathode and anode then accelerates these electrons, initiating localized Townsend avalanches throughout the volume. In GM mode this produces uniform pulses, which are subsequently shaped by preamplifiers and counted by discriminators.  


\begin{figure}[t!]  
\begin{center}  
\includegraphics[width=140mm]{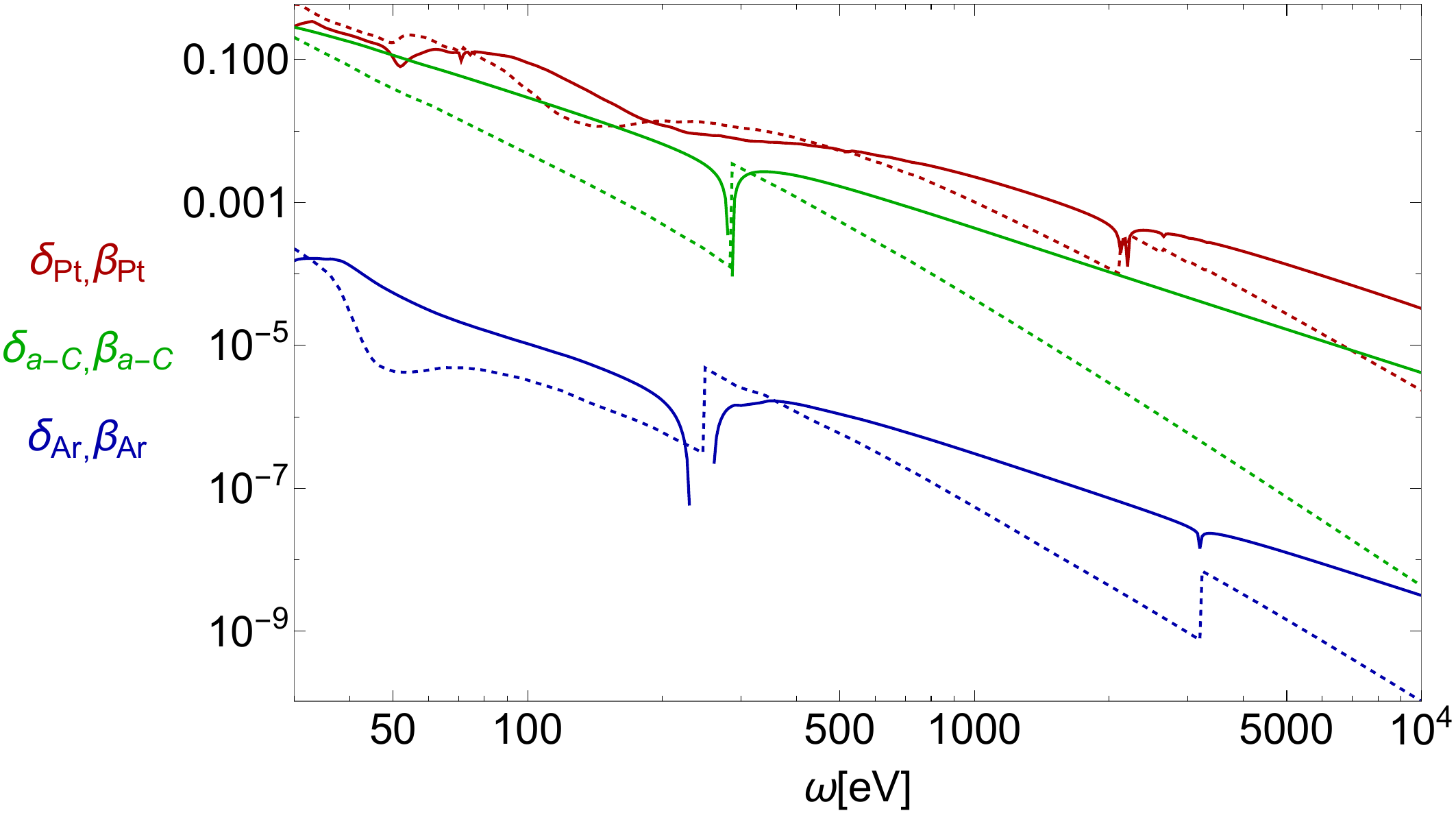}
\end{center}
\caption{
$\delta$ (solid lines) and $\beta$ (dashed lines) in the complex refractive indices of Pt (red), alphamos-C (green), and Ar (blue), from top to bottom.  
I assume densities of 21.5, 2.0, and 0.00166 $\rm g/cm^3$ for Pt, alphamos-C, and Ar, respectively. Values are estimated from \cite{henke1993x}.}
\label{fig:n}
\end{figure}

In practice, the GM counter has an effective energy response starting from a few keV, limited by the thickness ($\sim$5-10\,$\mu$m) of the mica or polyimide entrance window for ordinary photons.  
However, when searching for dark-photon-like states, the counter should retain sensitivity down to the sub-keV range, since dark photons traverse the entrance window essentially without attenuation.  
Unlike ordinary photons, dark photons are therefore not subject to the usual low-energy cutoff of 20-30 keV relevant for X-rays.

In what follows I consider only interactions of dark-photon-like states with Ar inside the GM counter.  
This is a conservative assumption, because dark photons could also interact with the entrance window to produce signals, which I do not include.

The first ionization energy of Ar is $\sim 16$\,eV, well below the keV scale.  
A dark-photon-like state interacts with Ar with probability \cite{An:2013yua, An:2014twa, Sakurai:2022cki, Yin:2025awb}
\begin{equation}
\laq{det}
P_{\rm det}^{\rm eff} =  
\chi^2 \times \frac{ m^4}{\left|m^2-\Pi_{\rm Ar}\right|^2}  
\times \frac{-\Im \Pi_{\rm Ar}}{2\omega} \times L_{\rm det} \,,
\end{equation}
where the middle factor $\tfrac{ m^4}{|m^2-\Pi_{\rm Ar}|^2}$ gives a suppression when $m\ll |\Pi_{\rm Ar}|$.  
The factor $-\Im \Pi_{\rm Ar}/(2\omega)$ denotes the linear attenuation coefficient (inverse mean free path).  
Finally, $L_{\rm det}\approx $10 cm corresponds to the effective length of Ar inside the detector.
Once the reaction happens with $\gtrsim \rm keV$ energies,  $> \O(10)$ pairs of initial electrons are produced. 
Then the Townsend avalanches occur. For the GM region, the enhancement of the number of the electron is $10^{8-10}$~\cite{Knoll:2010xta}. Thus for one dark photon event, $10^{9-11}$ electrons are emitted, which are well above the detection efficiency. 

Thus I arrive at the final formula for the detection probability per electron traversing the undulator,
\begin{equation}
N_{\rm det}=\sum_{\epsilon^{\pm}}\int \frac{d^3 \vec k^{(2)}_{\rm mir}}{(2\pi)^3}  
\left|{\cal M}\right|^2 \times P_{\rm det}(\vec{k}^{(2)}_{\rm mir}) \, .
\end{equation}
From this we obtain the effective photon--photon transmission rate,
\begin{equation}
P_{\gamma\to \gamma}^{\rm eff}\equiv \frac{N_{\rm det}}{N_{\gamma}} \,
\end{equation}
with $N_{\g}$ given in \Eq{photonum}. 
Assuming \Eq{photo}, the dark-photon signal rate is given by
\begin{equation}
P_{\gamma\to \gamma}^{\rm eff}\,\dot{n}_{\gamma}\, ,
\end{equation}
and I set the exclusion limit by requiring
\begin{equation}
P_{\gamma\to \gamma}^{\rm eff}\,\dot{n}_{\gamma} < 100 \ {\rm min}^{-1} \, ,
\end{equation}
which corresponds to $1\,\mu \rm Sv/h$, about one order of magnitude larger than natural background radiation, at which conventional GM counters typically trigger an alert.\footnote{See also \url{https://www-pub.iaea.org/MTCD/Publications/PDF/EPR-First_Res-PDA/html/ti5.htm}.  
The threshold $1\,{\m \rm Sv/h}$ should be larger than the actual radiation-safety level adopted in facilities, and thus provides a conservative limit.}  

\begin{figure}[t!]
  \begin{center}  
    \includegraphics[width=145mm]{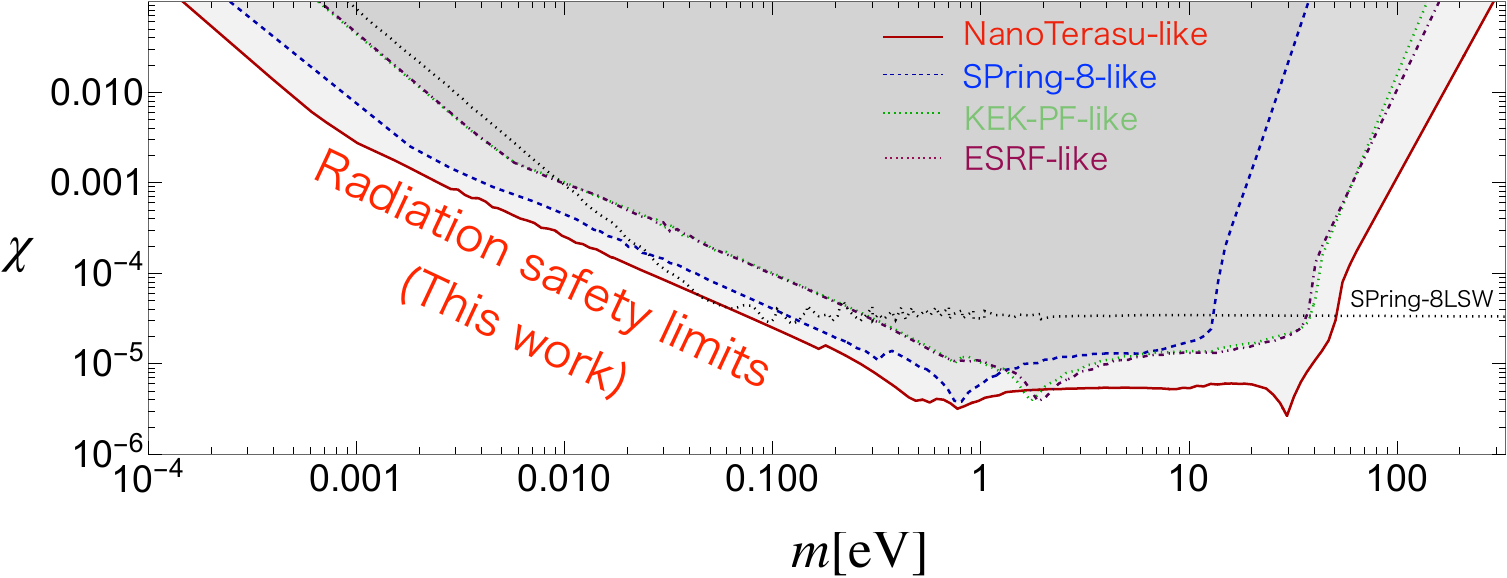}
  \end{center}
  \caption{Limits on the dark photon mixing parameter $\chi$ as a function of the dark photon mass $m$, derived from radiation safety considerations. 
 I set $L_{\rm det}=10$\,cm. 
  The strongest existing laboratory limit adopted from  \cite{Inada:2013tx} is shown by the dashed line.  }
  \label{fig:safety}
\end{figure}

The derived limits are shown in Fig.~\ref{fig:safety}, using the parameters in Table~\ref{tab:1}, which are chosen to mimic representative beam lines at NanoTerasu, SPring-8, KEK-PF, and ESRF for red solid, blue dashed, green dotted, and purple dot dashed lines, respectively.  
For comparison, the most stringent existing laboratory limit~\cite{Inada:2013tx} is also shown (dashed gray line).  
Since some technical details, such as mirror coatings, are not always available in the literature, I adopt conventional choices.  

At low masses ($m \lesssim \EV$), the sensitivity loss is mainly due to the suppression of the mixing through repeated Ar interactions. 
Depending on the energy, a narrow resonance can appear when $m$ is close to the plasma mass (see \Eq{det}), because at relatively high energies $\beta$ can be suppressed compared to $\delta$, as shown in Fig.~\ref{fig:n}.  
Around the eV scale, the derived limits can even surpass existing laboratory bounds.  
At even lower masses ($m \lesssim \rm meV$), the inefficient photon-dark photon oscillation further suppress the sensitivity. 

At higher masses, sensitivity decreases because undulators cease to directly produce dark photons ($m < m_1$), and straight transmission through mirrors is suppressed by the dispersion relation ($m < m_3$).  
The cutoff $m_3$ is controlled by the mirror angle $\theta_{\rm mir}$: larger $\theta_{\rm mir}$ allows more momentum modes of $f_\gamma$ to contribute to $N_{\rm mir}$ and  extends the sensitivity to higher masses.  
For example, with $\theta_{\rm mir}=2^\circ$ (amorphous carbon coating), the resonant region related to the plasma mass of carbon can even be reached.  

In practice, synchrotron facilities employ a variety of beam lines with different energies and mirror configurations, so the limits presented here should be regarded as representative examples.  
If radiation-safety monitoring were systematically applied across all beam lines, a much broader set of constraints could be obtained.  
\begin{table}[htbp]
\centering
\begin{tabular}{lcccccc}
\hline
$\gamma$ &
  $K$ & $2\pi/k$ & $L k/(2\pi)$ & $2\omega_{\rm typ}$ & Mirror coating & $\theta_{\rm mir}$ \\
\hline
 6000 & 
 1 & 30\,mm & 33  & $1.4$\,keV & Amorphous carbon &  $2^\circ$ \\
16000 & 
 1 & 40\,mm & 125 & $8$\,keV   & Pt & $0.1^\circ$ \\
13000 & 
1 & 20\,mm & 75 & $10$\,keV & Pt & $0.1^\circ$ \\
12000 & 
 0.5 & 20\,mm & 73 & $14$\,keV & Pt & $0.1^\circ$  \\
\hline
\end{tabular}
\caption{Representative parameters used in deriving the limits. 
Here $\omega_{\rm typ}\equiv k \gamma^2/(1+K^2)$ denotes the typical energy of the fundamental harmonic.  
The factor of 2 is included to indicate the on-axis value, following standard convention.  
The rows approximately mimic selected beamlines in NanoTerasu, SPring-8, KEK-PF(-AR), and ESRF (from top to bottom). Here I consider helical undulators, for simplicity, $\kappa=1$ and $\phi=0$.}
\label{tab:1}
\end{table}

\section{Conclusions}

In this work, I have shown that radiation-safety monitoring at synchrotron radiation facilities can be exploited as a novel method to search for dark photons.  
By analyzing the response of a simple Geiger-M\"uller (GM) counter, I derived new limits on the kinetic mixing parameter $\chi$, which in some mass ranges are competitive with or even stronger than existing laboratory bounds.  
Although the setups considered here are not identical to actual experimental conditions, the estimates are conservative---for instance, I neglect secondary photon-like states as well as the dark photon interaction with the entrance window of the counter, adopt a relatively large threshold of $1\,\m$Sv/h to set the limit, and assume only a few representative beamlines.  
Therefore, the derived bounds should be regarded as realistic, provided that radiation safety has indeed been monitored in front of the beam lines.  

In deriving the limits, I have found that the detector response to dark photons differs significantly from a na\"{i}ve $\chi^2$ scaling.  
Depending on the operating principle of the detector and the mass, both suppressions and enhancements of the detection rate can occur.  
This feature is in fact generic, and must be carefully taken into account when designing or interpreting experiments.  

This paper also demonstrates the potential impact of a ``parasitic experiment’’ at synchrotron radiation facilities~\cite{Yin:2024rjb}.  
So far, I have assumed a rather crude photon detector, namely a GM counter.  
It would be important to use a more precise X-ray detector to perform the real-time parasitic experiment, which could improve the limits significantly~\cite{Yin:2025awb}.

\section*{Acknowledgments}
W.Y. thank the useful discussions with Junya Yoshida, and  Toshio Namba. 
This work is supported by JSPS KAKENHI Grant Numbers  22K14029, 22H01215 and Tokyo Metropolitan University Grant for Young Researchers (Selective Research Fund).

\appendix 

\section{Photon wave packet}

By neglecting the dark photon mass, the wave packet of visible photons emitted from the undulator can be written as~\cite{Jackson:1998nia, Yin:2024rjb, Yin:2025awb}
\begin{align}
f_{\gamma}(k_\gamma) \simeq  
i \,\frac{e \chi K \pi}{\sqrt{2\omega}\,2k\gamma}
\left[
k \left( \kappa e^{-i\phi}\,\epsilon^x + i\,\epsilon^y \right)
+ \epsilon^z \left( \kappa e^{-i\phi} k_{\gamma}^x + i k_{\gamma}^y \right)\beta^z
\right]
\,\delta_{L}\!\left(\Delta k_\gamma - k\beta^z\right),
\end{align}
where $\epsilon^i$ denote polarization basis vectors, for which I only consider the transverse modes, and 
$
\Delta q \equiv q^{0}-\beta^{z}q^{z},
$
with $\beta^z$ being the electron motion in the beam direction ($z$-direction) and the $\g$ the corresponding Lorentz factor. 
The time averaged longitudinal velocity of the electron can be approximated as
\begin{equation}
\beta^z \simeq 1-\frac{1}{2\gamma^2}\left(1+\frac{1+\kappa^2}{2}\,K^2\right),
\end{equation}
where the correction of $\O(K^2)$ to the conventional $1/\g$ expansion is because the electron wiggling.

The finite-width delta function is defined as
\begin{equation}
\pi\delta_{X}(q)\equiv\int_{0}^{X}\!dx\,e^{iqx}
=\frac{e^{iqX}-1}{iq},
\end{equation}
which reduces to the usual Dirac delta in the $X\to\infty$ limit. 
In the numerical simulations, we introduce an exponential regulator,
\begin{equation}
\pi\delta_{L}(q)=\frac{e^{iqL}-1}{q}\;\longrightarrow\;
\frac{e^{iqL}-1}{q}\,e^{-\frac{\epsilon\,|q|L}{2\pi}},
\end{equation}
in order to suppress spurious enhancements at $k_{\gamma'} \gg \gamma^{2}k$ and $\theta\sim1$ (see Ref.\,\cite{Yin:2025awb} for further discussion).

\bibliography{radiation-safety.bib} 

\end{document}